\documentclass[amsmath,amssymb,preprint,showpacs]{revtex4}

\usepackage{graphicx}
\usepackage[usenames]{color}

\newcommand{\sub}[1]{_\textrm{#1}}

\newcommand{\micron}{$\mu$m}

\begin{document}

\title{Pressure-induced quenching of the charge-density-wave state observed by x-ray diffraction}

\author{A.~Sacchetti$^1$, C.L.~Condron$^2$, S.N.~Gvasaliya$^3$, F.~Pfuner$^1$, M.~Lavagnini$^1$, M.~Baldini$^4$, M.F.~Toney$^2$, M.~Merlini$^5$, M.~Hanfland$^5$, J.~Mesot$^3$, J.-H. Chu$^6$, I.R.~Fisher$^6$, P.~Postorino$^4$, and L.~Degiorgi$^1$}

\affiliation{$^1$Laboratorium f\"ur Festk\"orperphysik,
ETH-Z\"urich, CH-8093 Z\"urich, Switzerland. \\
$^2$Stanford Synchrotron Radiation Laboratory, Stanford Linear Accelerator Center, Menlo Park, CA 94025, U.S.A.\\
$^3$Laboratory for Neutron Scattering, ETH Z\"urich and Paul Scherrer Institute, CH-5232 Villigen PSI, Switzerland.\\
$^4$CNR-INFM-Coherentia and Dipartimento di Fisica Universit\`a ``La
Sapienza'', P.le A. Moro 5, I-00185 Rome, Italy.\\
$^5$European Synchrotron Radiation Facility, Bo\^ite Postale 220, F-38043 Grenoble, France.\\
$^6$Geballe Laboratory for Advanced Materials and
Department of Applied Physics, Stanford University, Stanford,
CA 94305-4045, U.S.A.
}

\date{\today}

\begin{abstract}
We report an x-ray diffraction study on the charge-density-wave (CDW) LaTe$_3$ and CeTe$_3$ compounds as a function of pressure. 
We extract the lattice constants and the CDW modulation wave-vector, and provide direct evidence for a pressure-induced quenching of the CDW phase. We observe subtle differences between the chemical and mechanical compression of the lattice. We account for these with a scenario where the effective dimensionality in these CDW systems is dependent on the type of lattice compression and has a direct impact on the degree of Fermi surface nesting and on the strength of fluctuation effects.
\end{abstract}



\maketitle

Low dimensionality plays an important role in condensed-matter physics owing to the observation of intriguing phenomena such as the formation of spin- and charge-density-waves (CDW), as well as non Fermi-liquid behavior of the electronic properties in a variety of materials \cite{CDW,LeoBook}. A revival of interest in low-dimensional interacting electron gas systems, and in their wealth of astonishing properties, took place since the discovery of high-temperature superconductivity in the layered two-dimensional (2D) copper oxides. This furthermore led to the quest for prototype layered systems, allowing a systematic study of these phenomena. In this context, the rare-earth tri-tellurides ($R$Te$_3$, with $R$=La-Sm, Gd-Tm) were recently revisited and recognized as a paramount example of easily tunable 2D materials. Their crystal structure is weakly orthorhombic (pseudotetragonal) \cite{struct,LattConst,foot_struct} and is composed of corrugated $R_2$Te$_2$ slabs alternating with pairs of Te-layers, stacked along the (long) $b$ axis. The formation of the CDW condensate, hosted within Te-layers, is to a large extent driven by the nesting of the Fermi-surface (FS) \cite{Dimasi,Brouet}, which is thus gapped over a sizeable portion. Systematic x-ray diffraction (XRD) studies of the $R$Te$_3$ series revealed that the modulation vector $\vec{q} \approx (2/7) \vec{c}^*$ ($\vec{c}^*$ is the reciprocal lattice vector of the $c$ axis) is almost the same for every member of this family \cite{qCDW}, and that the CDW state is progressively suppressed as the lattice is chemically compressed (i.e., going from $R$=La to $R$=Tm) \cite{ICDW}. The transition temperature $T\sub{CDW}$ is 250~K for TmTe$_3$, increases gradually up to 410~K in SmTe$_3$ \cite{ICDW}, and is larger than 450~K in the tri-tellurides with lighter rare-earth elements ($R$=La-Nd) \cite{qCDW}. A corresponding strong reduction of the CDW gap with chemical pressure was then established on the basis of optical-spectroscopy experiments \cite{IR_PRB}. 
Subsequent light scattering experiments on the same series of compounds showed that the CDW gap reduction is accompanied by a progressive disappearance of the signal from the Raman active phonon modes \cite{Raman}. The same effects have been observed in the light rare-earth tri-tellurides under external hydrostatic pressure: infrared reflectivity experiments on CeTe$_3$ at high-pressures \cite{IR_PRL} as well as on the related LaTe$_2$ compound \cite{LaTe2_HP} revealed a pressure-induced reduction of the CDW gap, while the Raman-active phonons in LaTe$_3$ and CeTe$_3$ disappeared \cite{Raman}. 

In this Letter we present a high-pressure XRD diffraction study on LaTe$_3$ and CeTe$_3$, with the goal of monitoring the evolution of the CDW distortion with pressure. We establish a pressure-induced quenching of the CDW state and show that, while there is general equivalence between physical and chemical pressure, there are also important subtle distinctions. We speculate that this is due to differences in the effective dimensionality that derive from the chemical and physical lattice compression and from the resulting interplay between FS instabilities and fluctuation effects.

The LaTe$_3$ and CeTe$_3$ single-crystals were grown as described in Ref.~\onlinecite{Ru}. Small ($\approx 20\times 20 \times 10$~\micron$^3$) sample slabs were placed inside the hole (initial diameter 250~\micron) of a stainless steel gasket of a membrane driven diamond anvil cell (DAC, culet size 600~\micron) together with He as pressure transmitting medium and a $\sim$ 5~\micron~diameter ruby sphere for pressure calibration \cite{Mao}. Diffraction images were collected at the ID09A beamline of the ESRF with a monochromatic beam ($\lambda=0.413 $~\AA) focussed to ~$ 20\times 20$~\micron$^2$ using a MAR345 image plate detector. During exposure the pressure cells were rotated around the $\phi$-axis. Here at $\phi$=0 the incident x-rays are along the $b$-axis and $\phi$ rotates the $b$ axis about the incoming beam direction. Total rotation ranges were 40$^\circ$ with standard 1.5 mm high diamond anvils and 60$^\circ$ with cells modified for Boehler-Almax anvils \cite{Boehler}. For the low temperature measurements the DAC was placed inside a He flow cryostat with the rotation range limited to $\pm 5^\circ$. These latter measurements were performed on coarse polycrystalline samples.

\begin{figure}[!tb]
\center
\includegraphics[width=0.85\columnwidth]{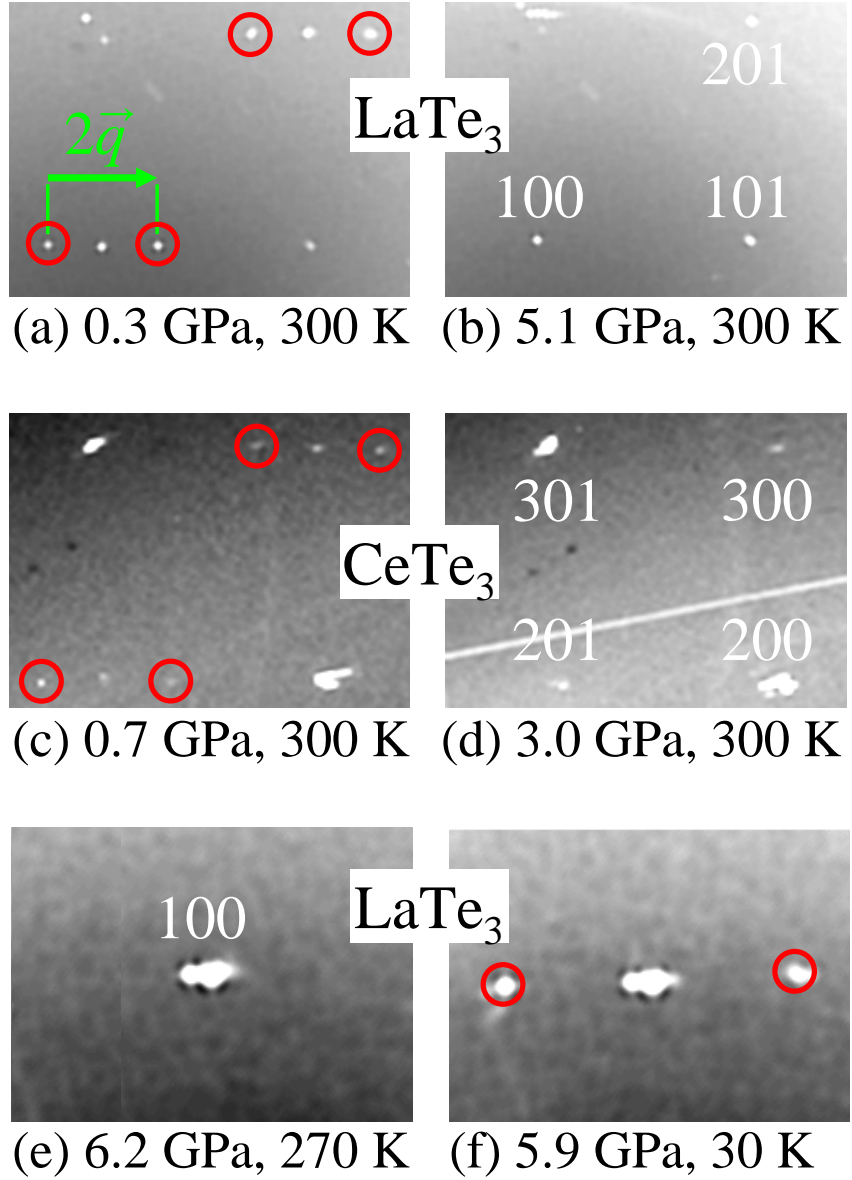}
\caption{ (color online) Selected 2D XRD patterns on single-crystals of LaTe$_3$ at 300~K and at 0.3~GPa (a) and 5.1~GPa (b), and of CeTe$_3$ at 300~K and at 0.7~GPa (c) and 3.0~GPa (d), and finally on polycrystalline LaTe$_3$ at 270~K and 6.2~GPa (e), and at 30~K and 5.9~GPa (f). Circles highlight the CDW satellite peaks. The modulation vector $\vec{q}$ is also shown in (a).}
\label{Fig_Data}
\end{figure}

XRD patterns were collected at room temperature as a function of pressure for LaTe$_3$ and CeTe$_3$ single crystal and as a function of temperature for a LaTe$_3$ polycrystal at $6.0\pm 0.2$~GPa. Representative areas of the diffraction patterns at selected pressures and temperatures are shown in Fig.~\ref{Fig_Data}.
At low-pressure and at 300~K, several Bragg-peaks in both LaTe$_3$ and CeTe$_3$ display pairs of satellites, which are due to the modulated CDW lattice-distortion \cite{ICDW}. Upon increasing pressure, the intensity of these satellite peaks is progressively reduced and they eventually disappear at high enough pressure (3 and 5~GPa in CeTe$_3$ and LaTe$_3$, respectively) as shown in Fig.~\ref{Fig_Data}(a-d). At $6.0\pm 0.2$~GPa the satellite peaks in LaTe$_3$ are recovered by cooling the polycrystalline specimen well below 300~K [see Fig.~\ref{Fig_Data}(e,f)]. This indicates that at this pressure the CDW transition occurs at a lower critical temperature $T\sub{CDW}$ (i.e., $<$ 300~K).

\begin{figure}[!t]
\center
\includegraphics[width=0.85\columnwidth]{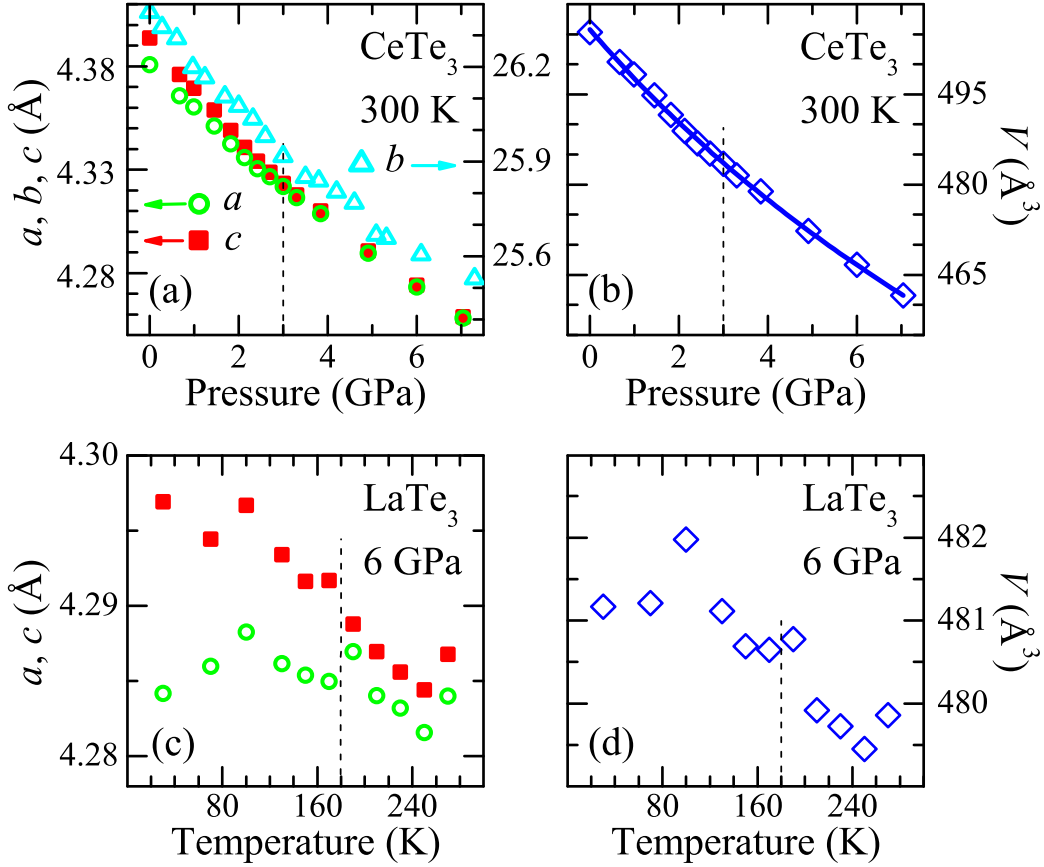}
\caption{(color online) Lattice parameters (left panels) and unit-cell volume (right panels) for CeTe$_3$ at 300~K as a function of pressure (a,b) and for LaTe$_3$ polycrystal at $6.0\pm 0.2$~GPa as a function of temperature (c,d). The solid line in (b) is the Birch-Murnaghan fit to the data \cite{Murnaghan}. Vertical dashed lines indicate the pressure and temperature where the CDW satellite peaks (Fig.~\ref{Fig_Data}) disappear.}
\label{Fig_abc}
\end{figure}

From the positions of selected Bragg peaks we obtained the lattice parameters and thus the unit-cell volume. The corresponding results for CeTe$_3$ under pressure at 300~K and for LaTe$_3$ as a function of temperature at $6.0\pm 0.2$~GPa are shown in Fig.~\ref{Fig_abc}.
The pressure experiment on LaTe$_3$ at room temperature provides similar results (not shown) as for CeTe$_3$. At ambient pressure the slight orthorhombic distortion of the unit-cell results in a small difference between the in-plane lattice parameters $a$ and $c$ \cite{foot_struct}. Upon increasing pressure this difference between the $a$ and $c$ axes decreases and both lattice constants become nearly indistinguishable above 3~GPa in CeTe$_3$ [Fig.~\ref{Fig_abc}(a)]. A similar effect is observed in LaTe$_3$ at 5~GPa and was previously reported for $R$Te$_3$ with $R$=Sm-Tm on crossing $T\sub{CDW}$ at ambient pressure \cite{ICDW}. The unit-cell volume [Fig.~\ref{Fig_abc}(b)] decreases smoothly with pressure and follows the Birch-Murnaghan equation of state \cite{Murnaghan}:
\begin{equation}
V(P) = V_0 \left(1+\frac{B'}{B_0}P \right)^{-1/B'}
\label{EqBM}
\end{equation}
with the bulk-modulus $B_0=59$~GPa and its pressure-derivative $B'=5.6$. The fitted $B_0$ value is in reasonable agreement with a previous estimate \cite{IR_PRL} from specific heat data \cite{SpecHeat} while $B'$ lies within the typical range of 4-8 \cite{Bprime}. A change in the temperature dependence of the $c$ and $a$ lattice constant is also observed when lowering the temperature below 180~K in LaTe$_3$ at 6~GPa [Fig.~\ref{Fig_abc}(c)]. While the scatter in the data is more pronounced than in the experiment as a function of pressure at fixed temperature, the tendency for $a$ and $c$ to diverge below 180 K is evident, like the weak discontinuity in the unit-cell volume [Fig.~\ref{Fig_abc}(d)]. The overall temperature dependence of the lattice parameters is very similar to that observed at ambient pressure in the heavy rare-earth tri-tellurides \cite{ICDW}, suggesting an analogous impact of both chemical and applied pressure on the structural properties of $R$Te$_3$.

\begin{figure}[!t]
\center
\includegraphics[width=0.85\columnwidth]{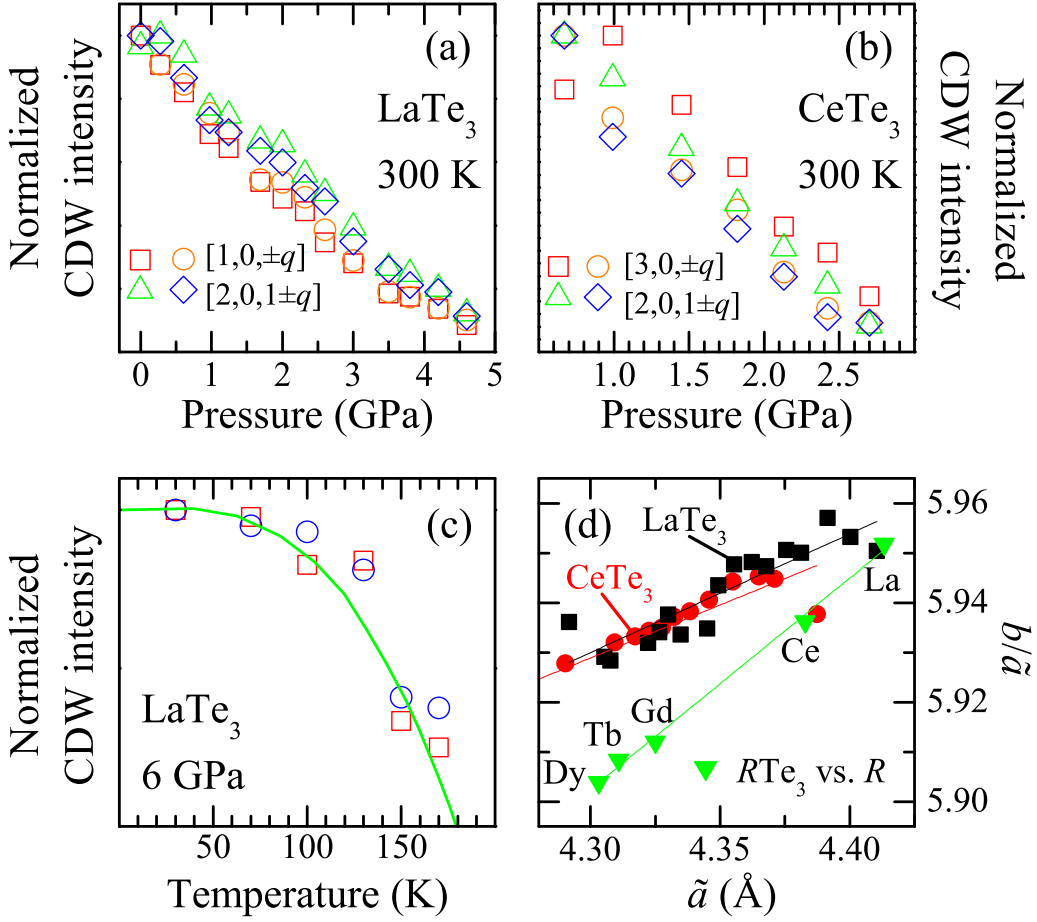}
\caption{(color online) Intensity of selected CDW satellite peaks normalized to a nearby Bragg-peak for LaTe$_3$ (a) and CeTe$_3$ (b) at 300~K as a function of pressure and for LaTe$_3$ at $6.0\pm 0.2$~GPa as a function of temperature, with the prediction from the BCS theory (c). (d) Ratio $b/\tilde{a}$ (see text) as a function of the average lattice constant $\tilde{a}$. Thin lines are linear interpolations to the data, as guide to the eyes.}
\label{Fig_Icdw}
\end{figure}
The most compelling result of our investigations is the observation that the integrated intensities of the CDW satellite peaks 
gradually decrease with increasing pressure at 300~K, and vanish at 5 and 2.8~GPa in LaTe$_3$ and CeTe$_3$, respectively [Fig.~\ref{Fig_Icdw}(a) and (b)]. This finding shows that the CDW state, observed at ambient pressure and 300~K in these compounds, is quenched by a moderate lattice compression of about 5\% of the volume. This is moreover consistent with the previously reported pressure-induced reduction of the CDW gap \cite{IR_PRL} and of the integrated intensities of the Raman-active phonon modes \cite{Raman}. The temperature dependence of the intensities of the CDW satellite peaks of LaTe$_3$ at 6~GPa [Fig.~\ref{Fig_Icdw}(c)]  is also similar to that of the heavy rare-earth tri-tellurides at ambient pressure \cite{ICDW}, supporting again a qualitative equivalence between chemical and applied pressure in order to achieve the lattice compression. We note that the satellite intensities are consistent with the BCS behavior \cite{BCS} with $T\sub{CDW}=180$~K, bearing a striking similarity with results on prototype 1D systems \cite{CDW}.  

\begin{figure}[!t]
\center
\includegraphics[width=0.70\columnwidth]{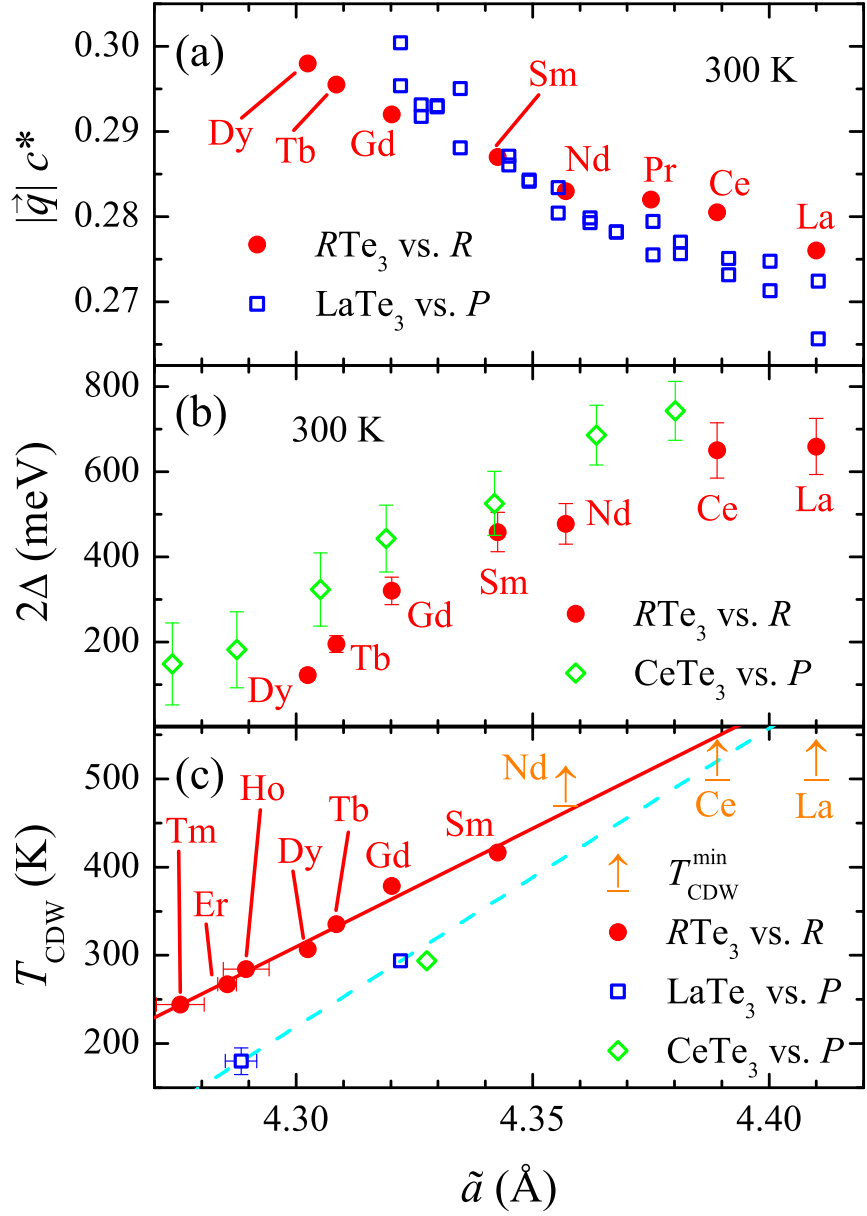}
\caption{(color online) (a) Magnitude of the CDW wave-vector $\vec{q}$ and  (b) of the CDW gap $2\Delta$ at 300~K for the $R$Te$_3$ series, and for LaTe$_3$ and CeTe$_3$ under applied pressure as a function of the average lattice constant $\tilde{a}$ (see text), respectively. Chemical-pressure data in (a) are from Ref.~\onlinecite{qCDW} and the $2\Delta$ values in (b) are from Refs.~\onlinecite{IR_PRB} and \onlinecite{IR_PRL} for $R$Te$_3$ and CeTe$_3$, respectively. 
 (c) CDW-transition temperature $T\sub{CDW}$ as a function of $\tilde{a}$ for the $R$Te$_3$ series \cite{ICDW,qCDW} 
 and for LaTe$_3$ and CeTe$_3$ at high pressures. The solid and dashed lines in (c) are linear fits to the data.}
\label{Fig_Merit}
\end{figure}

To further test the extent of equivalence between chemical and applied pressure, we first define the average in-plane lattice parameter $\tilde{a}=(a+c)/2$, which is related to the Te-Te distance within the Te-layers. $\tilde{a}$ can be considered as a common variable for several quantities measured upon compressing the lattice. We then plot in Fig. 3(d) the quantity $b/\tilde{a}$, which is representative of the anisotropy ratio between the inter- and intra-plane Te-Te distances, for the $R$Te$_3$ series \cite{LattConst,ICDW} and for LaTe$_3$ and CeTe$_3$ under applied pressure using the data from Fig. 2(a). Despite some scatter in the data there is a clear trend, showing a significantly more pronounced decrease of   $b/\tilde{a}$ in the chemical series than in the applied pressure experiment. The $\tilde{a}$-dependence of the CDW modulation wave-vector $|\vec{q}|$, extracted from the positions of the satellite peaks (Fig.~\ref{Fig_Data}), for the $R$Te$_3$ series at 300~K \cite{qCDW} and for LaTe$_3$ as a function of pressure is shown in Fig.~\ref{Fig_Merit}(a). The magnitude of $|\vec{q}|$ monotonously increases with increasing pressure. The similarity of $|\vec{q}|$ with chemical and applied pressure is suggestive of an equivalent modification of the FS and thus of its nesting properties \cite{Brouet,IR_PRB}, upon compressing the lattice. The $\tilde{a}$-dependence of the CDW gap $2\Delta$ (averaged over the FS) for the $R$Te$_3$ series at 300~K \cite{IR_PRB} and for CeTe$_3$ under pressure \cite{IR_PRL} is shown in Fig.~\ref{Fig_Merit}(b). The comparison here is much better than that reported in Ref.~\onlinecite{IR_PRL}, which was based on a crude estimation of $\tilde{a}(P)$. The two data sets follow the same trend and within the experimental error the $2\Delta$ values for both experiments are almost identical. From our data we can furthermore exclude that the CDW gap in the chemical $R$Te$_3$ series is larger than in CeTe$_3$ under pressure. Some educated guesses can be also advocated as far as the pressure dependence of the critical temperature $T\sub{CDW}$ is concerned. From our isothermal experiments at 300~K on LaTe$_3$ and CeTe$_3$ we obtain two points in the $\tilde{a}-T$ phase diagram and an additional data point is extracted from the isobaric experiment at $6.0\pm 0.2$~GPa on LaTe$_3$ [i.e., converting the pressure at which the satellite peaks disappear into a corresponding lattice constant (Fig.~\ref{Fig_abc})]. These data are then compared in Fig.~\ref{Fig_Merit}(c) with $T\sub{CDW}$ of the $R$Te$_3$ series measured at ambient pressure \cite{ICDW}. For $R$=La, Ce, and Nd only the lower limits of $T\sub{CDW}$ are known \cite{qCDW}.

While these findings confirm the overall qualitative equivalence between chemical and applied pressure, they show that a subtle difference exists between the two types of lattice compression. In contrast to our observations for $2\Delta$ [Fig.~\ref{Fig_Merit}(b)], we note that $T\sub{CDW}$ is systematically larger for the chemical pressure than for applied pressure. This is unexpected. We speculate that such a behavior is due to a difference in the effective dimensionality of the system when compressing the lattice chemically compared with applied pressure: specifically, the effective dimensionality is larger (more three dimensional) for chemical pressure. This intriguing possibility is consistent with the observation that the relative change of $b/\tilde{a}$ between LaTe$_3$ and DyTe$_3$ at ambient pressure is roughly a factor two larger than, for instance, in LaTe$_3$ between 0~GPa \cite{ICDW} and 5.5~GPa, for which $\tilde{a}$ is the same as in DyTe$_3$ at ambient pressure [Fig. 3(d)]. The lower effective dimensionality achieved by applied pressure implies stronger fluctuations and therefore a reduced transition temperature. At the same time, if the reduced dimensionality has an effect at all on the FS properties, this might lead to a better FS nesting and thus a possible larger (average) gap $2\Delta$. 
A detailed structural study devoted to the determination of the internal atomic coordinates in the unit-cell is required to confirm this speculative scenario. Nonetheless, a crude linear extrapolation of the $T\sub{CDW}$ data with chemical pressure and with applied pressure on LaTe$_3$ [Fig.~\ref{Fig_Merit}(c)] leads to an intersection at $\tilde{a} \approx 4.43$~\AA, which compares nicely with the zero-pressure value for LaTe$_3$ \cite{LattConst}.

In summary, we have reported a high-pressure XRD study on the CDW LaTe$_3$ and CeTe$_3$ compounds. The pressure dependence of the in-plane lattice parameters is consistent with a pressure-induced reduction of the $pseudo$-tetragonal phase, i.e. of the lattice distortion accompanying the formation of the CDW condensate. This is similar to what has been observed upon cooling across the CDW transition in LaTe$_3$ at high pressure (present work), as well as at ambient pressure in the heavy rare-earth tri-tellurides \cite{ICDW}. More striking evidence of the pressure-induced quenching of the CDW phase is provided by the intensities of the CDW satellite peaks, which tend to zero with increasing pressure. Such observations support ideas based on the equivalence between chemical and applied pressure in $R$Te$_3$, put forward in our previous work \cite{IR_PRB,Raman,IR_PRL}. Nevertheless, subtle differences between the two types of lattice compression were revealed. Those could be accounted by differences in effective dimensionality, and hence the impact of fluctuations and FS nesting, upon lattice compression achieved by either chemical substitution or hydrostatically. 

\begin{acknowledgments}
The authors wish to thank R. Monnier for fruitful
discussions. This work has been
supported by the Swiss National Foundation for the Scientific
Research as well as by the NCCR MaNEP pool and also by the Department of
Energy, Office of Basic Energy Sciences under contract
DE-AC02-76SF00515. Portions of this research were carried out at the Stanford Synchrotron Radiation Laboratory, a national user facility operated by Stanford University on behalf of the U.S. Department of Energy, Office of Basic Energy Sciences. 
\end{acknowledgments}


\begin{thebibliography}{99}

\bibitem{CDW} G. Gr\"uner, \emph{Density Waves in Solids}, Addison Wesley, Reading, MA (1994).

\bibitem{LeoBook} \emph{Strong Interactions in Low Dimensions}, Eds. D. Baeriswyl and L. Degiorgi, Kluwer Academic Publishers, Dordrecht (2004).

\bibitem{struct} B.K. Norling and H. Steinfink, \emph{Inorg. Chem.} \textbf{5}, 1488 (1966).

\bibitem{LattConst} P. Villars and L.D. Calvert, \emph{Pearson's Handbook of Crystallographic Data for Intermetallic Phases}, American Society for Metals, Metals Park, OH (1991).

\bibitem{foot_struct} The crystal structure of $R$Te$_3$ belongs to the $Cmcm$ space group \cite{struct}, which is orthorhombic. However, the in-plane lattice constants $a$ and $c$ differ by approximately ($c$-$a$)/$a$ $\sim 0.3\%$ and 0.5\% in CeTe$_3$ and LaTe$_3$, respectively (in the standard space group setting, the $b$-axis is perpendicular to the Te planes).

\bibitem{Dimasi} E. DiMasi \textit{et al.}, \emph{Phys. Rev. B} \textbf{52}, 14516 (1995).

\bibitem{Brouet} V. Brouet \emph{et al.}, \emph{Phys. Rev. Lett.} \textbf{93}, 126405 (2004).

\bibitem{qCDW} C.D. Malliakas and M.G. Kanatzidis, \emph{J. Am. Chem. Soc.}, \textbf{128}, 12612 (2006).

\bibitem{ICDW}  N. Ru \textit{et al.}, \textit{Phys. Rev. B} \textbf{77}, 035114 (2008).

\bibitem{IR_PRB} A. Sacchetti \textit{et al.}, \emph{Phys. Rev. B} \textbf{74}, 125115 (2006).

\bibitem{Raman} M. Lavagnini \textit{et al.}, \textit{cond-mat} 0806.1455 (2008), to be published in \emph{Phys. Rev. B} (in press).

\bibitem{IR_PRL} A. Sacchetti \textit{et al.}, \emph{Phys. Rev. Lett.} \textbf{98}, 026401 (2007).

\bibitem{LaTe2_HP} M. Lavagnini \textit{et al.}, \textit{Phys. Rev. B} \textbf{77}, 165132 (2008).

\bibitem{Ru} N. Ru and I.R. Fisher, \emph{Phys. Rev. B} \textbf{73}, 033101 (2006).

\bibitem{Mao} H.K. Mao, J. Xu, and P.M. Bell, \textit{J. Geophys. Res.} \textbf{91}, 4673 (1986).

\bibitem{Boehler} R. Boehler and K. De Hantsetters, \textit{High Press. Res.} \textbf{24}, 391 (2004).
 
\bibitem{Murnaghan} F.D. Murnaghan, \textit{P. Natl. Acad. Sci. U.S.A.} \textbf{30}, 244 (1944).

\bibitem{SpecHeat} K.Y. Shin \textit{et al.}, \emph{Phys. Rev. B} \textbf{72}, 085132 (2005).

\bibitem{Bprime} S. Jiuxun \textit{et al.}, \textit{J. Phys. Chem. Solids} \textbf{66}, 773 (2005).

\bibitem{BCS} M. Tinkham, {\em Introduction to superconductivity}, $2^{nd}$ Ed., McGraw-Hill, New York (1996).


\end{thebibliography}
\end{document}